\documentclass[sigconf]{acmart}

\usepackage[linesnumbered,ruled,vlined]{algorithm2e}
\setcopyright{none}
\renewcommand\footnotetextcopyrightpermission[1]{} 
\AtBeginDocument{%
  \providecommand\BibTeX{{%
    \normalfont B\kern-0.5em{\scshape i\kern-0.25em b}\kern-0.8em\TeX}}}

\begin{document}

\title{FEBR: Expert-Based Recommendation Framework for beneficial and personalized content}



\author{Mohamed Lechiakh}
\affiliation{%
  \institution{Mohammed VI Polytechnic University}
  \city{Ben Guerir}
  \country{Morocco}}
\email{mohamed.lechiakh@um6p.ma}

\author{Alexandre Maurer}
\affiliation{%
  \institution{Mohammed VI Polytechnic University}
  \city{Ben Guerir}
  \country{Morocco}}
\email{alexandre.maurer@um6p.ma}



\begin{abstract}
So far, most research on recommender systems focused on maintaining long-term user engagement and satisfaction, by promoting relevant and personalized content.
However, it is still very challenging to evaluate the quality and the reliability of this content. In this paper, we propose FEBR (Expert-Based Recommendation Framework), an apprenticeship learning framework to assess the quality of the recommended content on online platforms. The framework exploits the demonstrated trajectories of an expert (assumed to be reliable) in a recommendation evaluation environment, to recover an unknown  utility function. This function is used to learn an optimal policy describing the expert's behavior, which is then used in the framework to provide high-quality and personalized recommendations. We evaluate the performance of our solution through a user interest simulation environment (using RecSim). We simulate interactions under the aforementioned expert policy for videos recommendation, and compare its efficiency with standard recommendation methods. The results show that our approach provides a significant gain in terms of content quality, evaluated by experts and watched by users, while maintaining almost the same watch time as the baseline approaches.
\end{abstract}

\ccsdesc[500]{Information systems}
\ccsdesc[500]{Information systems~Recommender systems}
\ccsdesc[300]{Theory of computation~Apprenticeship learning}

\keywords{Recommender systems, apprenticeship learning, reinforcement learning, expert policy, maximum entropy}

\maketitle

\section{Introduction}

Recommender systems (RS) try to provide their users with content matching their interests and preferences. To do so, they use many different approaches: collaborative filtering, content-based approaches, hybrid approaches\dots \cite{8276172}. 
 Recent works on collaborative interactive and conversational RS showed promising methods to improve the relevance and the personalization  of recommendations~\cite{10.1145/2645710.2645753,10.1145/2939672.2939746, 10.1145/3397271.3401419,adomavicius2005toward}. They mainly focus on modeling complex user behaviors and dynamic user/item interactions. For this purpose, they use techniques derived from optimal control paradigm, preference elicitation learning, deep learning and natural language processing (NLP)~\cite{ 10.5555/3327546.3327641,zou2019reinforcement, 10.1145/3336191.3371801, 10.1145/3289600.3291604,10.1145/3292500.3332290}. Nowadays, recommendation platforms are often judged to be lacking transparency and accountability with their algorithmic recommendations, which have a tendency to capture user attention and make the platform more addictive. Thus, user values and principles are sometimes underestimated, and even intentionally suppressed, to fulfill the company's objectives. \footnote{For instance, according to the report published by \textit{New America}'s website~\cite{newAmerica}: YouTube has faced particular criticism for creating a ``rabbit hole'' effect, in which the algorithm delivers personalized recommendations that prompt users to consume harmful or radical content that they did not originally seek out.}
 In fact, aligning recommendations with human values is a complicated problem that requests (1) a good understanding of user behaviors and (2) optimizing the right metrics. Here, these metrics must be carefully designed to adapt to user  preferences and goals in a fair, accurate and ethical way (e.g., by reducing addictive, harmful, misleading and polarizing content).

Actually, finding beneficial information within mainstream RS is far from obvious. This problem becomes very challenging when trying to personalize the recommendation service, and it is especially prevalent with news and informative contents provided by those platforms. In the case of video RS like YouTube, the content quality must be discussed from many perspectives, which includes: the amount of expertise, authoritativeness and trustworthiness of the content creator; the validity and accuracy of the content itself; its ability to create and support good habits and behaviors; the user engagement and satisfaction towards this content; etc. These quality features characterize what we call a \emph{beneficial} personalized content. According to \textit{YouTube Official Blog} \cite{youtubeblog}, YouTube recently adopted the concept of social responsibility as a core value for the company. In this context, it defined new metrics for quality evaluation (namely ``user watch time'' and ``quality watch time''). In addition, it started recommending from sources that the company considers to be authoritative, and reducing suggestions of ``borderline'' videos. However, despite these efforts, YouTube's recommendation algorithm still suggests content containing misleading or false information, as well as conspiracy theories. Therefore, in such huge RS, high-quality content can be ``drowned'' among low-quality content, making it less visible to users.\footnote{For instance, in the context of the COVID-19 pandemic, countries found themselves fighting against a huge amount of misinformation, uploaded everyday on YouTube and social networks. This resulted in a considerable number of users being mislead with false information, inducing dangerous habits and behaviors. Such behaviors have likely increased the propagation rate of the virus in many countries.} 
Unfortunately, designing robust ML methods to recommend beneficial and engaging content seems hopeless in systems of large dimension, with billions of users and items. 

\paragraph{\textbf{Our contribution.}}

To address this problem in current recommendation environments, we propose \textbf{FEBR} (\textbf{F}ramework for \textbf{E}xpert-\textbf{B}ased \textbf{R}ecommendations), the first expert-based Apprenticeship Learning (AL)~\cite{10.1145/3054912} framework for news (video, articles\dots) RS. In our solution, we try to use relevant quality signals to identify beneficial and personalized content to be recommended to users. The novelty in our approach is in the way we measure this quality, and the guarantees that it presents in terms of importance, accuracy and reliability.
We derive our quality metrics from experts involved in the evaluation part of the framework, using a customized evaluation model. Personalization is essentially achieved by the classifier which matches the user state model to its closest expert state model (from the expert state dataset). Our simulation results show a consumption of high-quality content that remarkably exceeds baseline approaches, while maintaining a very close total watch time
(as shown in Figures~\ref{fig:wtime} and \ref{fig:ttq} of Section~\ref{sec_expres}). In addition, we implemented our solution as a configurable framework (using RecSim~\cite{ie2019recsim}), that can be used for simulation experiments related to our topic.

\paragraph{\textbf{Overview of our solution.}}
We took inspiration from the WeBuildAI framework~\cite{Lee2018WeBuildAIP}: a collective participatory framework that enables stakeholders to provide useful inputs for learning personalized models in order to create algorithmic policies. Our proposed framework FEBR is a three-part participatory system where the RS, experts and users collaborate between each other in Markov Decision Process (MDP) environments, to leverage expert knowledge for a better user experience. We assume that experts are reliable (we are aware that they can only evaluate a small fraction of the system's items). An intuitive idea would be to directly inject the set of evaluated videos as a ``ground truth'' of reliable quality content, that would be exploited by collaborative filtering techniques to improve the quality of recommended contents.
However, this approach is very limited, since it relies on the probability that a given evaluated video will likely be explored and recommended based on the similarities between user profiles (which would not be efficient in RS with large user and item spaces).
Therefore, we started from the assumption that a reliable expert's behavior across the recommended contents (for a given topic) is different from a layman's one.\footnote{For instance, a healthcare specialist, when reviewing videos about COVID-19 on YouTube, may carefully select these videos based on their titles, references and descriptions. Then, she can judge their quality, according to her expertise domain.}
Thus, if we could guide a user to mimic an expert's trajectory (i.e., her sequence of visited states within the recommendation environment) while respecting her own preferences, we would achieve high quality while maintaining high engagement. However, it seems to be difficult to exactly know the expert's intentions throughout her watching and evaluation session, since she is dealing with an interactive environment with a large action space. Therefore, it is almost impossible to capture an expert's behavior through a RL method that tries to optimize a given reward function: in this case, unlike classical problems, we cannot specify the reward. Indeed, the reward would depend on many observable and latent environment parameters, including the expert and system's states, and the complex dependencies and relationships this may involve.

Besides the difficulty of inferring the objective of the expert's behavior from a demonstration (i.e., an expert trajectory within a continuous session), we believe that, in general, in a partially observable MDP environment, her demonstration is actually a series of state-action operations that aim to maximize an unknown reward function, which reflects different aspects of the expert's intuitions, estimations, choices and evaluations.
Thus, our problem can be seen as an AL problem that uses Inverse Reinforcement Learning (IRL)~\cite{10.1145/3054912,Ng00algorithmsfor} to recover a complex reward function in an expert MDP environment (within an AL/IRL component).
This reward function will be used to generate the optimal expert policy, which will be consumed by the final user MDP environment (in the recommendation component of the same RS) to generate high-quality content.
We assume that this reward function can be formulated as a linear combination of unknown features describing the learning task.
We developed a dedicated expert MDP environment that uses a personalized evaluation mechanism within the response model. This enables experts to evaluate videos and, at the same time, to capture the main patterns of quality and engagement features delivered during this process (which could help to efficiently learn the reward function).
However, the expert demonstrations cannot always be optimal (i.e., some demonstrations may include bad content selections, and lead to a low average quality), which may scramble the learning process and make it difficult to converge to a stable reward function.
To overcome this ambiguity, we use the Maximum Entropy (MaxEnt) principle with IRL \cite{ziebart2008maximum} to select the best distribution over trajectories, leading to a maximum reward value.
Finally, we use the \emph{value iteration} \cite{Sutton1998} algorithm to optimize the policy under the recovered reward function, which will be exploited by a classifier to build our recommender agent for the end-user MDP environment. It is crucial to mention that our approach recommends items to the user according to the policy action (slate) learned on a similar expert state. 
This expert state is selected (based on a classification algorithm) for its similarity with the user state.
Also, we do not discuss the problem of disagreement between experts, since they all maintain their own evaluations and the system converges to a unique optimal (expert) policy.

\paragraph{\textbf{Organization of the paper.}}
In Section~\ref{sec_rw}, we present related works and the background.
In Section~\ref{sec_prelim}, we state preliminary definitions.
In Section~\ref{sec_fram}, we describe our proposed framework.
In Section~\ref{sec_expset}, we describe the design and setting of our experimentation.
In Section~\ref{sec_expres}, we describe and comment our experimental results.
We conclude and discuss future works in Section~\ref{sec_conc}. 

\section{Related works}
\label{sec_rw}
In this section, we discuss some important works around AL and IRL, and how they are (or can be) related to RS.
Many works on AL have been applied in the robotic domain, and most of them consider some supervised learning techniques to learn a mapping from the states to the action space. However, such methods are not effective in highly dynamic environments, and can even lead the agent (apprentice) to choose catastrophic actions. Abbeel and Ng.~\shortcite{conf/icml/PieterN04} provided a short overview of this literature. In their paper, they supposed that an expert is trying to optimize for an unknown reward function, that can be expressed as a linear combination of known features in a finite MDP environment, in order to find a policy that performs as well as the expert. However, assuming that the reward function is a linear combination of hand-selected features may be unrealistic.

Neu et al.~\shortcite{DBLP:journals/corr/abs-1206-5264} proposed another approach based on a gradient algorithm that combines the supervised learning versions of AL with Abbeel and Ng's method to learn a cost function by tuning the reward parameters, and using RL to find the optimal policy.
This approach is demonstrated for unknown reward features, but both of them use IRL, which regularly calls RL algorithms to solve MDP problems.
They further suppose finite MDP settings, which is far from convenient in large interactive systems, like many of today's RS.

A few works studied the issue of learning a non-linear reward function using some probabilistic reasoning about stochastic expert behaviors with Gaussian processes. They managed to recover both the reward and the hyper-parameters of a kernel function that describes the structure of the reward~\cite{NIPS2011_4420}.
In addition, many policies can be a solution for a given reward, and many reward functions can model a given set of demonstrations.
To tackle this problem in small dynamic spaces, Ziebart et al.~\shortcite{ziebart2008maximum} proposed a maximum entropy probabilistic framework using IRL, which aims to choose a trajectory distribution with maximum entropy that matches the expected expert reward. This approach is further extended by Wulfmeier et al.~\shortcite{DBLP:journals/corr/WulfmeierOP15} using a deep neural network to learn the unknown reward features. Here, the maximum entropy principle is used to optimize the weights of the neural network.

IRL algorithms use the entire trajectories of the learned task instead of an independent state-action operation to learn a similar expert policy by maximizing an unknown reward function.
Contrary to the approach used by Zhao et al.~\shortcite{DBLP:journals/corr/abs-1906-11462}, which considers a recommendation session as multiple independent state-action pairs, Chen et al.~\shortcite{chen2019generative} used a Generative Adversarial Imitation Learning approach~\cite{NIPS2016_6391} to learn a user behavior model and the corresponding reward function, which are used to develop a combinatorial recommendation policy using RL.
They considered that a sequence of state-action pairs is a whole trajectory, such that posterior actions could be influenced by prior actions.
However, both approaches succeed in improving the user's long-term engagement, regardless of the quality of the content they provide.

With an approach closer to our contribution, Massimo and Ricci~\shortcite{10.1145/3240323.3240392} proposed a tourism recommender system that builds a user behavioral model based on the policies learned from clustering the user's trajectories, then solved an IRL problem to derive the reward function of each cluster.
Their clustering approach could be seen as a projection of ours in a specific application domain using other tools, where the experts are exactly the same users that build ``correct'' trajectories by highlighting relevant points of interest from their own visits.

Existing works so far rely on the behavior of regular users, or try to inject some ``expert knowledge'' (e.g. guidelines, good practice\dots) in the system for improving the accuracy of predictions. For instance, some works~\shortcite{10.1145/1571941.1572033,4216978} used a small dataset of expert ratings collected from a reduced number of experts (identified by specific techniques like ``domain authority'' reputation-like score) to effectively predict the ratings of a large population. These approaches, based on the principle of ``wisdom of the few'', have only ensured the relevance of recommendations with the user preferences.  
To the best of our knowledge, this paper is the first to leverage the behavior of a set of selected experts to deliver beneficial, high-quality and personalized content.

\section{Preliminaries}
\label{sec_prelim}

\paragraph{\textbf{RL definitions.}} A MDP is defined in forward reinforcement learning (RL) by the tuple (\textit{S,A,T,D,R,$\gamma$}), where \textit{S} is the state space and \textit{A} is the action space.
$T : S \times A \times S \mapsto [0, 1]$ is the state transition 
function. $D$ is the initial-state distribution, from which the initial state $s_0 \in S$ is drawn. $R : S \times A \times S \mapsto  \mathbb{R}$ is the reward function. $\gamma$ is the discount rate. A stationary stochastic policy (we simply say ``policy'' from now) $\pi:S\times A\mapsto [0,1]$ corresponds to a recommendation strategy which returns the probability of taking an action given a user state $s_t$ at timestamp $t$. A policy is called deterministic if it results in a single action for any state of $S$.
Thus, for any policy $\pi$, the value of a state $s\in S$ w.r.t. the initial state distribution $s_0\sim D$ is given by $V^{\pi}(s) = E[\sum_{t}^{ \infty}\gamma^{t}R(s_t,a_t)|D,T]$, where $(s_t,a_t)_{t\geq0}$ is the sequence of state-actions pairs generated by executing policy $\pi$. A policy maximizing the value function such that $V^\star(s)=max_{\pi}V^{\pi}(s)$ for each $s\in S$ is called an optimal policy.

In general, a typical interactive RL-based RS executes an action $a_t=\{i_t\in I\}$ through recommending a slate of items $i_t$ (e.g., videos, commercial products) to a user who provides a feedback $f_t\in F_u$ (e.g., skipping, clicking and other reactions) at the $t^{\text{th}}$ interaction. Then, the RS recommends the next slate ($a_{t+1}$) until the user leaves the platform (end of session). Here, $I$ (resp. $F_u$) is the set of candidate items for recommendations (resp. set of possible feedbacks). We model a user MDP state by: $s_t=(user,a_1,f_1,...,a_{t-1},f_{t-1})$; then, a trajectory is represented by $\zeta_u=(s_0,a_0, r_0,...,s_t,a_t,r_t,...,s_T)$, where $r_t\in \mathbb{R}$ is a reward associated with the user's feedback. In the case of forward RL, this reward is
a function to be maximized by the recommender agent, which will derive its optimal policy.

\paragraph{\textbf{AL/IRL definitions.}} Algorithms for AL problems take as input a Markov decision process (MDP) with an unknown reward function (MDP$\setminus R$).\footnote{More generally, we can consider a partially observable MDP (POMDP$\setminus R$), if we are dealing with noisy states.} In a expert MDP$\setminus R$ environment, we observe an expert trajectory $\zeta_e$ through a sequence of state-action pairs $\zeta_e = (s_t,a_t)_{t\in\mathbb{N}}$ such that $s_t=(expert,a_1,f_1,...,a_{t-1},f_{t-1})$. Here, the expert set feedback $F_e$ would have particular reactions (evaluation features).  We assume that the expert behaves according to an optimal policy $\pi_{e}$, which is assumed to maximize an unknown reward function $R=R^*$ (typically, it is common to make some assumptions on the structure of the reward function, such as assuming a linear model). As discussed in the related works, there are many IRL algorithms and approaches~\cite{Ng00algorithmsfor,ziebart2008maximum} that could be used to find the optimized reward function. This function is recovered by a joint-iterative improvement and evaluation process using its associated policy, which is derived using some RL methods.

\section{Proposed framework}
\label{sec_fram}

\begin{figure}[h]
  \centering
  \includegraphics[width=0.9\columnwidth]{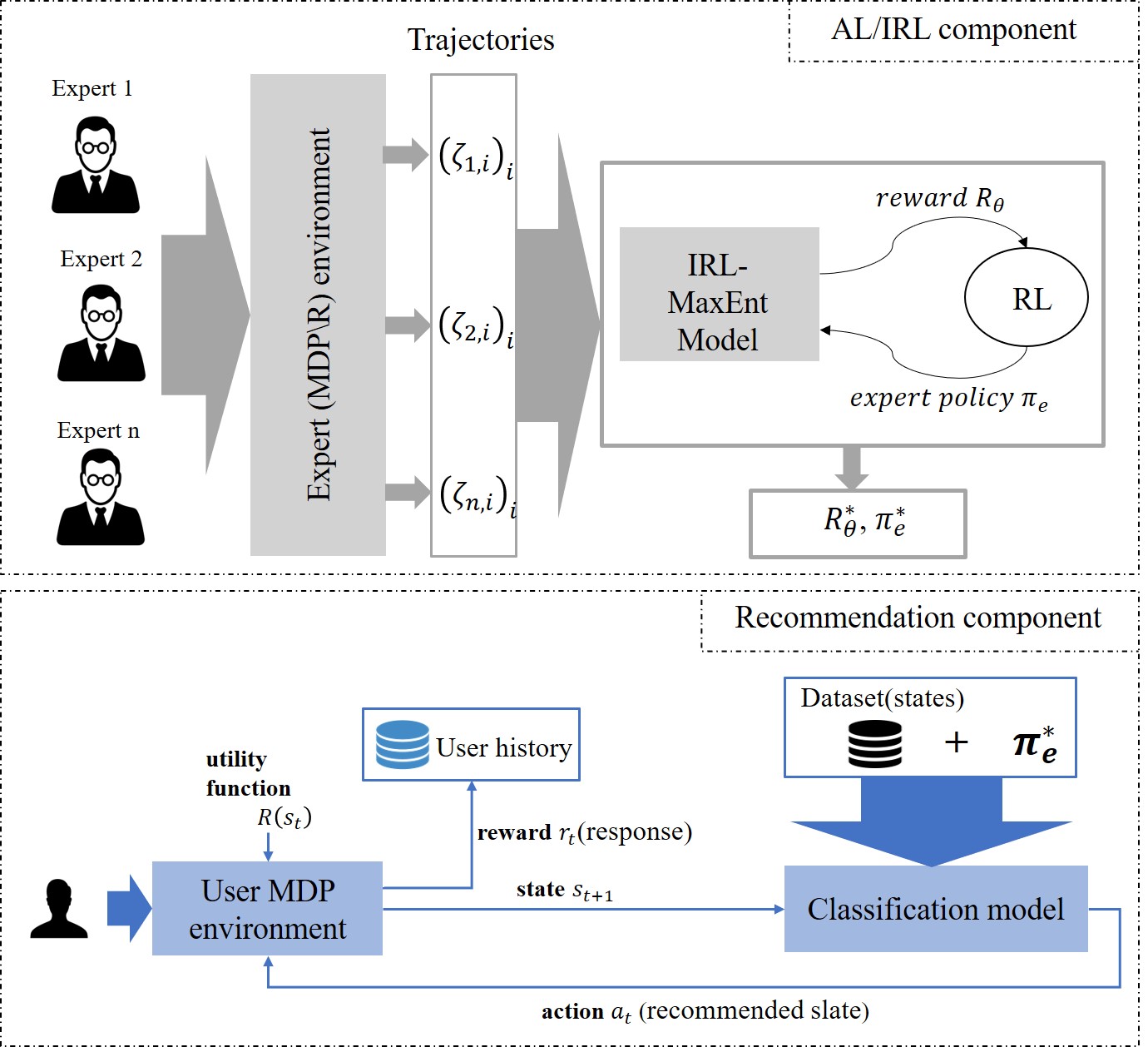}
  \caption{FEBR model framework.}
    \Description{FEBR model framework}
  \label{fig:fram}
\end{figure}

As shown in Figure~\ref{fig:fram}, the framework is composed of the AL/IRL component and the recommendation component. This model framework is a centralized version of our approach, where the experts collaborate to learn the same policy. Furthermore, we suppose that the framework operates sequentially through three stages. The first stage concerns the generated trajectories from all experts participating in the system. The second stage concerns the learning of the expert policy by the MaxEnt model. The third stage concerns the recommendation process using the expert classification model in a convenient user environment. In this section, we describe our framework in the context of video recommendation systems, but the approach remains valid for other types of instructive and informative recommended contents.

\subsection{Expert module statement}

\paragraph{\textbf{Expert MDP$\setminus$R environment.}} Following the AL/IRL component in Figure~\ref{fig:fram}, the mean objective of the expert environment is to capture the experts' behavior through their watching and evaluation process. The experts are assumed to be reliable and to act in a professional and ethical way. In this case, we developed a finite-state MDP$\setminus R$ environment to model the system behavior in front of experts' demonstrations. Then, for an expert session, the recommendation task is defined as the sequential interactions between a recommender system (agent) and the expert environment. It uses a MDP$\setminus R = (S, A, T, \gamma, D)$ environment to model it, where\dots
\begin{itemize}

\item $\mathbf{S}$ reflects the expert state space that refers to her interest distribution over topics, and the information about her watching history (e.g., watched videos and the interactions they involved: clicks, ratings using evaluation features, engagement rate\dots).\footnote{In our model framework, we propose a configurable evaluation model (see details in \ref{eval_model}) in which we can define the quality metrics to evaluate the total quality of each video. This model can be specified w.r.t. the nature of the service and the constraints of the system.}

\item $\mathbf{A}$ is the set of actions that contains all possible recommendation slates of size $k$. Basically, each action is a slate of videos that the system recommends to the expert. In a simple version, this slate can contain at least one item, or the null item $\perp$ when nothing is selected.\footnote{Note that having only one recommended item per slate may not allow a useful expert interaction, since we rely on its ability to distinguish relevant contents among recommended items.}

\item $\mathbf{R}$ is the reward function when the system takes an action $a_t$ from a state $s_t$. Basically, this reward is unknown, and it is influenced by (1) the expert choice model (when she selects a video from the slate) and (2) the feedback that she gave from this video during watch time. 
\end{itemize}

The set of state transition probabilities $\mathbf{T}$ and the initial-state distribution  $\mathbf{D}$ are determined by the environment models. Finally, let $\mathbf{\gamma} \in [0,1)$ be the discount factor.

For a given expert, instantiating this environment at timestamp $t$ gives us the state model: $s_m^t=$ [expert state, response state, video state], which forms the time steps of expert trajectories. These trajectories enable to construct the state dataset $\mathcal{D}_s$ (used by the recommendation component) with the same state models structure (see \ref{dataset} for more details). Thus, $s_m^t$ forms, with previous states (in the same trajectory), the general state model $s_t=(s_m^1,...,s_m^t)$ of the expert environment inside the AL/IRL component.

\paragraph{\textbf{IRL-MaxEnt model.}} For our contribution, we use the principle of maximum entropy \cite{ziebart2008maximum} built on Abbel and Ng.~\shortcite{conf/icml/PieterN04}'s approach. The true reward function $R^*$ is unknown, but we assume that we can linearly define its structure as $R^*(s) = \theta^T \phi_{s}$, 
where  $\phi_s:S\mapsto \mathbb{R}^k$ is a vector of $k$ features over the set of states, describing the expert's utility for visiting each state. This function is parameterized by some reward weights $\theta$. Then, given the path counts $\phi_{\zeta}=\sum_{s_t\in \zeta}^{} \phi_{s_t}$, the reward value of a trajectory $\zeta$ is:
\begin{equation}
reward(\theta_\zeta) = \theta^\top \phi_{\zeta} = \sum_{s_t\in \zeta}^{}R^*(s_t)
\label{eq1}
\end{equation}

Given $m$ trajectories $\Tilde{\zeta_i}$ extracted from the expert's behavior, its expected empirical feature count is  $\Tilde{\phi} = \frac{1}{m}\sum_{i}^{} \phi_{\Tilde{\zeta_i}}$. The idea here is to find a probability distribution $P$ over the entire class of possible trajectories to ensure the following equality:
\begin{equation}
\Tilde{\phi} = \sum_{i}^{}P(\zeta_i|\theta)\phi_{\zeta_i}
\label{eq2}
\end{equation}

Abbel and Ng~\shortcite{conf/icml/PieterN04} demonstrated that the matching feature expectations (Equation~\ref{eq2}) between an observed policy (of the expert) and a learner's behavior (of the RL agent) is both necessary and sufficient to achieve the same performance as this policy, when the expert is solving a MDP with a reward function linear in those features (Equation~\ref{eq1}). However, the matching problem of Equation~\ref{eq2} can result in many reward functions that correspond to the same optimal policy, and many policies can lead to the same empirical feature counts. Therefore, this approach may be ambiguous, especially in the case of sub-optimal demonstrations, like in our case, where the expert's behavior is often not a perfect one (since she is performing an evaluation work where optimal performance is not always expected). In our case, a mixture of policies is required to satisfy Equation~\ref{eq2}.

In the context of our contribution, we would like to fix a distribution that results in the minimum bias over its paths. In other terms, we look for a stochastic policy that only considers the constrained feature matching property (Equation~\ref{eq2}) and does not add any additional preferences on the set of paths (other than the ones implied by this constraint). Thus, Ziebart et al.~\shortcite{ziebart2008maximum} proposed to use the Maximum Entropy principle to resolve this problem. The selected distribution yielding higher total reward is an exponentially preferable choice, and it is parameterized by the reward weights $\theta$. Then, it turns out to be an optimization problem:

\begin{equation}
\theta^*=\arg\max_{\theta}\sum_{i}^{}log P(\zeta_i|\theta)
\end{equation}

We then maximize the likelihood of observing the expert demonstrations under the maximum entropy distribution $P$. This can be solved by a gradient optimization method:
\begin{equation}
\Delta L(\theta) = \Tilde{\phi} - \sum_{i}^{}P(\zeta_i|\theta)\phi_{\zeta_i} = \Tilde{\phi} - \sum_{i}^{}D(s)\phi_s
\end{equation}

where $D(s)$ is called \emph{expected state visitation frequency} of state $s$, and represents the probability of being in such a state. Ziebart et al.~\shortcite{ziebart2008maximum} presented a dynamic programming algorithm to compute $D(s)$. 
Within the gradient optimization loop,  we use a \emph{value iteration} ~\cite{Sutton1998} RL algorithm to derive the expert policy $\pi_e$, whose reward function is learned and optimized by the IRL-MaxEnt model. After the convergence of IRL-MaxEnt algorithm, we obtain the expert optimal policy $\pi^*_e$.

\subsection{FEBR recommendation component}
The recommendation component presented in Figure~\ref{fig:fram} is designed to host the end-user environment. Basically, it is an exploitation and evaluation component of the AL/IRL component. We feed this component with the state dataset $\mathcal{D}_s$ (presented in \ref{dataset}) and $\pi^*_e$ (learned by AL/IRL). The trick here is that we use $\mathcal{D}_s$ and $\pi^*_e$ to build a classification model (introduced in \ref{classi_model} to be used as a RL agent in a standard MDP recommendation (user) environment with a utility function $R(s_t)$ -- e.g., measures of quality, watch time, user long-term engagement\dots). The choice and the design of the classification model is a parameter of the system, and could be further studied as an independent problem. Note that, in this component, we use a well-known utility (reward) function, to avoid confusion with the AL/IRL methodology. In the proposed framework, the document model of such an environment must include a quality computation mechanism to rank items (videos) in the document corpus of each state (see \ref{quality_model} for more details). Furthermore, the rewards $r_t$ are used to construct a user's history dataset, which can be exploited by the AL/IRL component to improve the evaluation process of videos. They can also be used by the user environment for content personalization, and to evaluate the performance of the recommendation process by building potential new metrics.

\section{Experimental design and setting}
\label{sec_expset}
For the evaluation of our framework, we choose to perform simulated experiments (as previously done for several RL-based RS~\cite{10.1007/978-3-030-49663-0_29, 10.1145/1282100.1282114, 10.1145/359784.359829}), since most public datasets are irrelevant and not designed for evaluating multi-step user-recommender interactions (static datasets).
Furthermore, live experiments would be very expensive. Thus, we used the RecSim~\cite{ie2019recsim} platform to develop the simulation environment of our framework, in the case of a video recommendation platform.

\subsection{Simulation environment}
\label{simu_env}
RecSim provides the necessary tools and models to create RL simulation components for RS. We note that it is possible to re-implement our solution using another RL recommendation tool that respects the structure and the framework design of RecSim. In the following, we present an overview of the models used in our simulation, but we highly recommend to refer to the implementation project and to the RecSim paper~\cite{ie2019recsim} for more details about each model (response, choice, transition\dots) and the hypotheses that have been considered. Our source code is available here: \\\url{https://github.com/FEBR-rec/ExpertDrivenRec}

\paragraph{\textbf{Expert environment.}} In the AL/IRL component of our framework, we developed a POMDP$\setminus R$ expert environment with an IRL agent\footnote{The observation includes noisy information about the expert's reaction to the content, and potential clues about the user's latent state.} to model the regular behavior of experts (selecting, watching and evaluating videos). This environment is used to generate trajectories with a variable number of steps (defined by the consumption of time budget of the session). At each time step $t$, the IRL agent observes the expert state $s_t$ and chooses the slate to recommend using a (possibly stochastic) policy according to the unknown $R$. For generality, we do not specify the RL algorithm used by this agent to learn its policy, but we assume that it executes a stationary policy. Besides, we used a document model that implements a simple evaluation mechanism based on the quality features of a video: pedagogy, accuracy, importance and entertainment (see see \ref{eval_model} for more details).
\label{user_env}

\paragraph{\textbf{User environment.}} To evaluate the performance of $\pi^*_e$ in the recommendation component, we used the \textit{interest evolution} environment provided by RecSim, which mainly implements the user simulation environment described in \cite{ie2019reinforcement}. For this environment, we have to specify a reward function to be optimized by the agent. Then, we specify this reward according to the evaluation metrics (defined in \ref{sec_metrics}) that we will choose for the evaluation of our framework, which are quality and watch time. In addition, we propose a simple classification model (see \ref{classi_model}) to recommend, for each user state, the best slate specified by the learned $\pi^*_e$ based on a similar expert state. Ideally, to build this classification model, we could have used a ML classification algorithm trained on large $\mathcal{D}_s$, but the size of our generated dataset $\mathcal{D}_s$ is not sufficient to learn an accurate model.

For both expert and user environment, we used the choice and quality models that are introduced respectively in \ref{choice_model} and \ref{quality_model}.

\subsection{Choice model.}
\label{choice_model}
In our experiments, we use the conditional logit model~\cite{10.5555/355102} as the main choice model for both expert (AL/IRL component) and end-user (recommendation component) simulation environments. We recommend using conditional choice functions of the form $P(j|l)=u(x_{i,j})/\sum_{k\in l}^{}u(x_{i,k})$, in which an individual $i$ selects item $j$ from slate $l$ with an unnormalized probability $u(x_{i,j})$, where $u$ is a function of the user-item feature vector $x_{i,j}$. The conditional logit (and multinomial logit)~\cite{10.5555/355102} model is a common instance of this general format, which effectively captures an individual's behavior within her interaction environments. Cascade choice models~\cite{pmlr-v37-kveton15} are effective at capturing position bias through ordered lists of recommendations. Thus, cascade models could be interesting for experiments with large slate size and a large video corpus. Overall, it is highly recommended to use the same choice model for both user and expert environments, to ensure the consistency of the system requests and the accuracy of the results.

\subsection{Quality model.}
\label{quality_model}

We recall that our main objective is to build a system that recommends beneficial contents. In general, we call $q(s_t)$ the quality of the clicked (and probably watched) videos related to a given state $s_t$. The closer $q(s_t)$ is to 1 (resp. -1), the better (resp. worst) quality we get; 0 corresponds to a neuter quality. 
Typically, as for many practical RS, FEBR  scores/ranks candidate videos using a DNN with both user/expert context and video features as input, by optimizing a combination
of several objectives (e.g., clicks, expected engagement, satisfaction and other factors). These scores are often considered to describe the quality of videos in the RS. In FEBR RS, we generally use the same technique to update the value of $q(s_t)$, which is initially set to follow a uniform distribution $U(-1,1)$. However, $q(s_t)$ may be differently updated inside (1) the expert AL/IRL environment when providing evaluations after watching a video by an expert (see \ref{eval_model} for more details), and (2) the user environment of the recommendation component when the user clicked a video that has already been  evaluated by an expert. In the second case, $q(s_t)$ has the value from this evaluation.

\subsection{Configurable evaluation model}
\label{eval_model}
Similarly to an information retrieval system, RS should help users to make quality searches in order to achieve relevant quality recommendations. However, constructing efficient quality metrics for the evaluation of RS is still a very challenging problem. Within the expert environment, we propose a simple model to evaluate the quality of videos based on four quality criteria: \textit{Eval} = \{pedagogy, accuracy, importance, entertainment\}. This is a non exhaustive list, which could be modified and adapted depending on the evaluation procedure of the studied problem. Then, these features are initially set to 0, and $q(s_t) \sim U(-1,1)$ ($q(s_t)$ is the state quality introduced in \ref{quality_model}). Then, when a video is evaluated by an expert (for instance, by adjusting the features related to a session's state $s_t$ with a dedicated expert Web interface), an average quality score is calculated: $s_v=(1/4)\sum_{f_i\in Eval}^{}f_i$, and the video is marked as evaluated by this expert. In this case, we update  $q(s_t)$ by adding $s_v\times f$ ($f\in[0.1]$ is the expert quality factor that represents her amount of expertise in the topic of the video; it is a metadata of the expert's profile).

\subsection{State dataset $\mathcal{D}_s$}

\label{dataset}
$\mathcal{D}_s$ is constructed from expert trajectories where each entry contains a total description of the interaction information generated by the expert's environment. These entries have this model state form: [expert ID, expert state, response state, video state], where (1)
the \textbf{expert state} contains the interest distribution vector over topics $e_s\in \mathbb{R}^n$, where $n$ is the number of topics;
(2) the \textbf{response state} contains the expert behavior on the recommended slate (clicked video, watch time, values of evaluation metrics, engagement rate, new estimated quality); and
(3) the \textbf{video state} contains videos of the corpus sampled for this expert environment's state. For each video, we store its topic, length and quality.

\subsection{Classification model}

Basically, the implemented version of this classification model in FEBR is a similarity search algorithm (alg. \ref{alg1}) using Euclidean distance $d(u,v)= \lVert u-v \rVert$. This algorithm tries to match a user state model $s_u$ to her closest expert state model $s_e\in \mathcal{D}_s$ based on similarity between user/expert and corpus features. $s_u$ is defined by the inputs $u_i$ and $u_c$, such that the vector $u_i$ describes the user interest distribution over topics and $u_c$ contains video descriptors (topic, length, score or latent quality). Then, for each $s_e$, we extract the same vectors as $u_i$ and $u_c$ for the case of experts, which are the variables $e_s$ and $e_c$. $th1$ and $th2$ are respectively the \textit{interest margin} and the \textit{corpus margin}. The values chosen for the inputs $th1$ and $th2$ depend on the size of $\mathcal{D}_s$ (smaller values make the classifier more accurate, but highly selective). When an expert state $s_e\in \mathcal{D}_s$ is determined to match the user inputs, the underlying action (the output $slate_s$) learned by the expert policy $\pi^*_e$ is proposed. Otherwise, a random slate from the current user corpus $u_i$ is proposed.

\label{classi_model}
\begin{algorithm}
\SetAlgoLined
\textbf{Inputs: }$u_i$, $u_c$, th1, th2 \; 
\textbf{Output: } policy's action $slate_s$ \; 
\For{$s_e \in \mathcal{D}_s$}{
  \If{$d(u_i,e_s)\leq th1$ and $d(u_c,e_c)\leq th2$}{
   $slate_s \leftarrow$  Apply $\pi^*_e$ to $s_e$ to extract the related policy's action \;
   \Return $slate_s$
  }
 }
 \Return random $slate_s$ from $u_c$

 \caption{Classification of user latent states}
 \label{alg1}
\end{algorithm}
\subsection{Baseline recommendation approaches}
\label{baseline}

To evaluate the performance of FEBR, we consider the quality $q(s_t)$ of recommended videos that have been chosen by users. 
As explained in \ref{quality_model}, recommendation quality of most (if not all) existing approaches is essentially measured by how much the user appreciates their recommendations, and how successful they are in keeping the user engaged with the content of their system. On the opposite, with our solution, we try to first ensure the recommendation of a correct ``beneficial'' content, and then, to ensure that this content is as personalized as possible. To better compare our approach (that we now call \emph{RecFEBR}) with other methods, we make the following (conservative) assumption.

\textbf{Assumption.} We consider that the notion of quality used by other baseline methods (that  do not use expert evaluations,
namely \emph{RecFSQ}, \emph{RecPCTR} and \emph{RecBandit})
is the same as \emph{RecFEBR}, even if their measuring models are totally different.
\footnote{Actually, this is an unfair comparison (for our approach), because it could involve the comparison of a video quality estimated to reflect the attractiveness of an element for a given user with a quality that characterizes the content itself, assessed by an expert. Therefore, even though we get similar results, this is a very conservative estimate for FEBR.}

We compare \emph{RecFEBR} to\dots
\begin{itemize}
    \item \emph{RecFSQ}, a standard, full slate, Q-learning based approach.\footnote{This approach was first discussed by Sunehag et al.~\shortcite{DBLP:journals/corr/SunehagEDZVC15}, and then implemented in RecSim~\cite{ie2019recsim} as a baseline non-decomposed Q-learning method to test new recommender agents.} \emph{RecFSQ}'s RL algorithm recommends slates are based on a deep Q-Network (DQN) agent. This algorithm converges quickly in systems of small dimensions, to a policy that offers high average quality of recommendations compared to the SlateQ decomposed method \cite{ie2019reinforcement}.
    
    \item \emph{RecPCTR}, which implements a myopic greedy agent that recommends slates with the highest pCTR (predicted click-through rate) items. Basically, this agent receives observations of the true user and document states, because it assumes knowledge of the true underlying choice model.
    
    \item \emph{RecBandit}, which uses a bandit algorithm to recommend items with the highest UCBs of topic affinities~\cite{10.1023/A:1013689704352}. This agent exploits observations of user's past responses for each topic, without assuming any knowledge of their related user's affinity. Within the same best topic, the agent picks documents with high quality scores.
    
    \item \emph{RecNaive}, which is based on a random agent \cite{ie2019reinforcement} that recommends random videos from the corpus of videos with high expert ratings, which best match the current user context.
\end{itemize}

 We emphasize that the approaches \emph{RecFSQ}, \emph{RecPCTR} and \emph{RecBandit} do not consider quality values that are assigned by experts. Instead, they consider an inherent quality feature that represents the topic-independent attractiveness to the average user. Then, they reflect the general performance of popular RL-based RS using different techniques and powerful RL algorithms. They essentially try to maximize a reward function (using quality and watch time as engagement metrics) which leads to optimized learned actions. \emph{RecNaive} is useful to test the efficiency of our approach for personalizing high-quality content for each user. Note that all approaches, including ours (\emph{RecFEBR}), use the same user environment, which is introduced in Section~\ref{user_env}. 

\subsection{Evaluation metrics}
\label{sec_metrics}
We first investigate the efficiency of our solution for recommending positive contents. Therefore, we analyze the quality of watched videos that are recommended based on the learned expert policy $\pi^*_e$. We define $\mathcal{S}_e$ as the set of user states in which the system recommends a slate following $\pi^*_e$ (states that are successfully matched by the classification model). For the comparison with baseline approaches, we propose to evaluate the performance of FEBR by measuring the average quality of the watched videos, as well as the total watch time, as an engagement metric showing how much the user was interested in the proposed content.
 Then, for a full user session $\mathcal{S}_u$  of length $l$ (where $\mathcal{S}_e \subseteq \mathcal{S}_u$), we use the following evaluation metrics:
\begin{itemize}
    \item Average expert-based quality: $Q_e=\frac{1}{|\mathcal{S}_e|}\sum_{s_t\in \mathcal{S}_e}^{}q(s_t)$
    \item Average total quality: $Q_T=\frac{1}{l}\sum_{t=0}^{l}q(s_t)$
    \item Total watch time of selected videos: $W_T$
\end{itemize}

\subsection{Setting} 

We consider a global system of 8 categories of video topics. We assume that each video belongs to a single category (note that this is for experiments simplification: it is still possible to consider the assignment to many categories). The simulation is performed upon a large set of videos (order of magnitude $10^5$); for each state, the system retrieves a small corpus candidate of size 5 that best matches the user context, then recommends a slate of size 2 (small values are chosen for technical constraints). 
\paragraph{\textbf{AL/IRL training.}}
We consider a sub-system of 10 experts. For each expert, we generate 100 trajectories, each one with at most 20 steps. We fix $\gamma=0.5$ for the \emph{value iteration} RL algorithm. In such a complex system, determining the best value of $\gamma$ is not obvious, since it is further constrained by the performance of a learned policy based on unknown (and insignificant) rewards. Basically, values close from 1 would result in a better learning, reducing the number of iterations required to optimize sub-optimal policies; but it would also result in a much higher convergence time. For the \textit{IRL-MaxEnt} algorithm, we set the number of training iterations to 10000. This simulation runs on CPU using a server machine equipped with an Intel Xeon Gold 6152/2.1GHz, and 384G DDR4 RAM.

\paragraph{\textbf{Recommendation component.}} According to the explanation in \ref{classi_model} for the classification model, we set $th1=0.5$ and $th2=0.1$. We recommend using this setting for a generated dataset $\mathcal{D}_s$ of less than 50000 states (the larger it gets, the better it is to use small values). We simulate 3000 independent user sessions. We run this simulation on a desktop computer equipped with an Intel i7-8550U CPU, GTX 720ti, and 8G DDR4 RAM.

\section{Experimental results}
\label{sec_expres}

In this section, we propose to discuss the efficiency of our approach by analyzing the quality metrics measured by the user simulation environment (recommendation component). Then, we evaluate our framework, and compare it to the baseline approaches introduced in section \ref{baseline}. 

\subsection{Analysis of content quality metrics}

\begin{figure}[h]
  \centering
  \includegraphics[width=0.6\columnwidth]{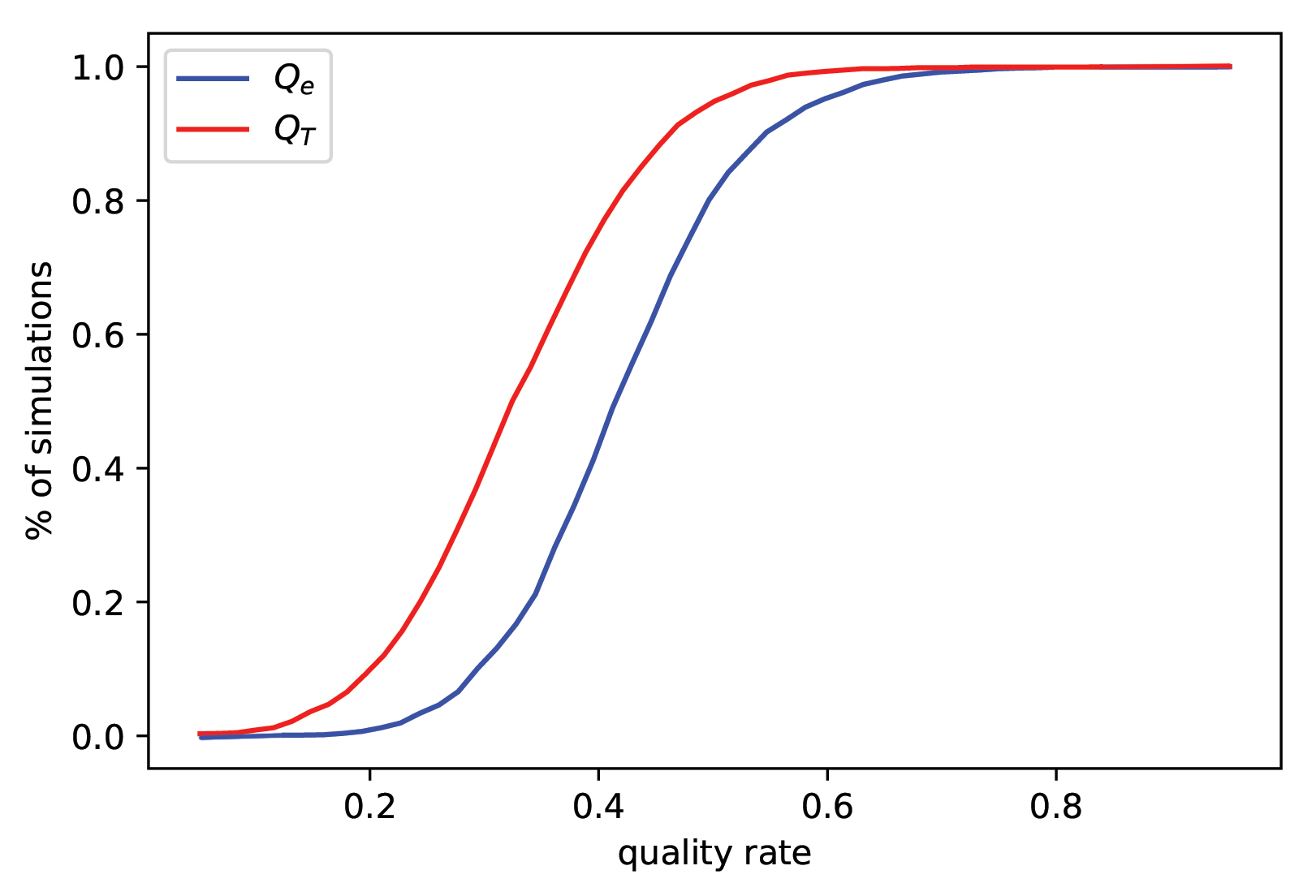}
  \caption{Cumulative distribution of $Q_T$ and $Q_e$ for 3000 simulations.}
  	\label{fig:febr_q}
\end{figure}
Figure~\ref{fig:febr_q} shows that FEBR efficiently leads to positive quality consumption ($Q_e$) when following $\pi^*_e$. More precisely, the $Q_T$ plot presents the average total quality which is achieved within all watched videos for a given session. If the classification model succeeds to match a user's state $s_t$ to its closest expert's state (i.e., $s_t\in \mathcal{S}_e$), then, the recommended slate will be the underlying expert slate fixed by $\pi^*_e$. In this case, this results in the average expert-based quality $Q_e$. Otherwise, for unsuccessful matching operations, the framework recommends arbitrary slates (with random quality/score) from the current user corpus. Therefore, we can write $Q_T=Q_e + Q'$, such that $Q'$ is the quality achieved through those arbitrary recommendations.
We notice that $Q_e$ always results in increasing positive values.
Thus, arbitrary recommendations are likely responsible of providing content of low quality ($Q'<0$) that decreases the value of $Q_T$.
The values of $Q_e$ show that indeed, this simulation setting produced a good expert-like policy $\pi^*_e$. On the other hand, the values of $Q'$ reflect some weakness of the recommendation component. This weakness could be caused by many factors such as the small size of $\mathcal{D}_s$, inaccurate classification model and the cold start problem. Nevertheless, these problems remain inevitable in large and dynamic systems. 

\subsection{Evaluation and comparison}

\begin{figure}[h]
\begin{minipage}[b]{0.48\linewidth}
\centering
\includegraphics[width=\textwidth]{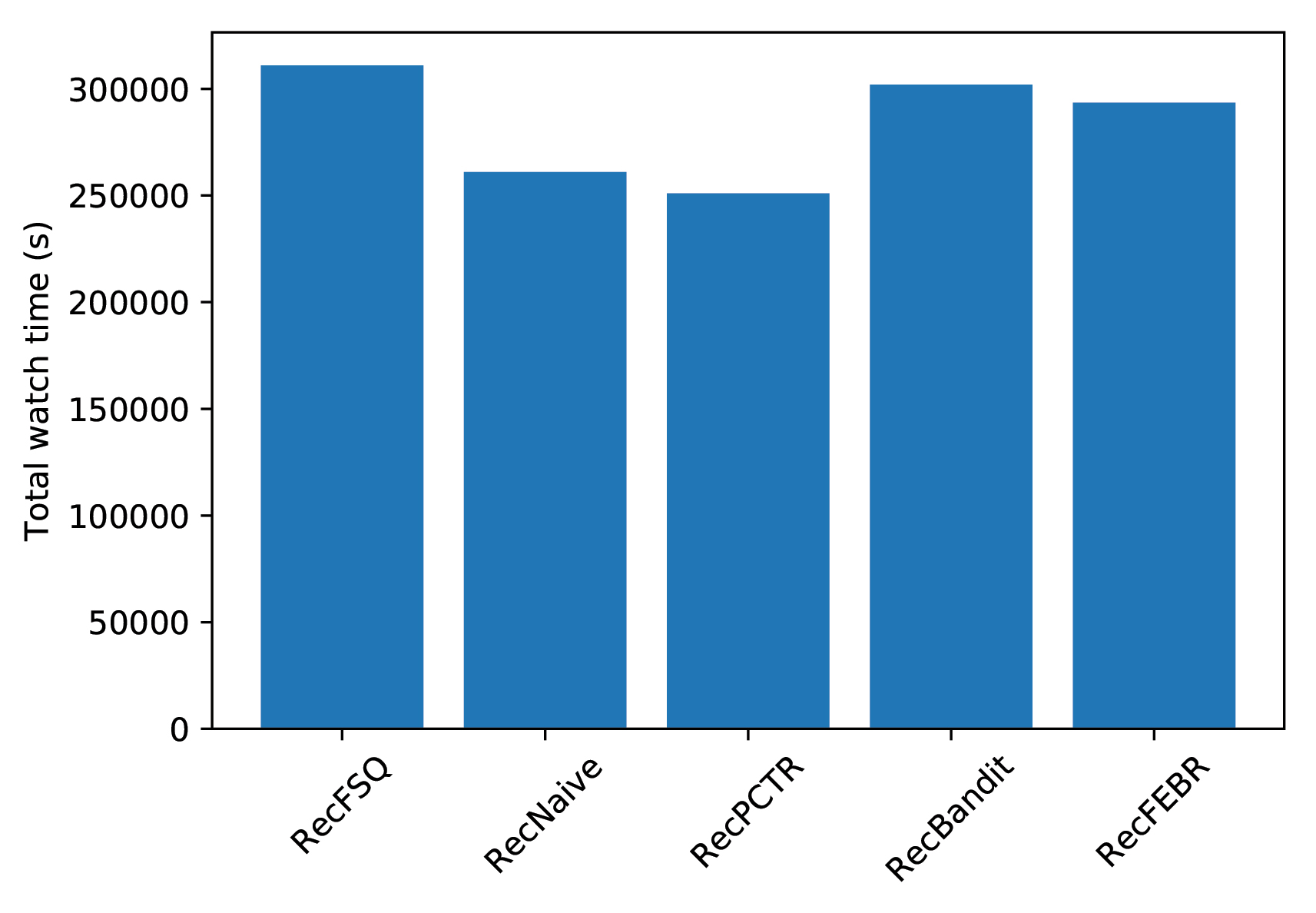}
\caption{Comparison of the total watch time $W_ T$ of baseline approaches and \textit{RecFEBR}.}
\label{fig:wtime}
\end{minipage}
\hspace{0.1cm}
\begin{minipage}[b]{0.48\linewidth}
\centering
\includegraphics[width=\textwidth]{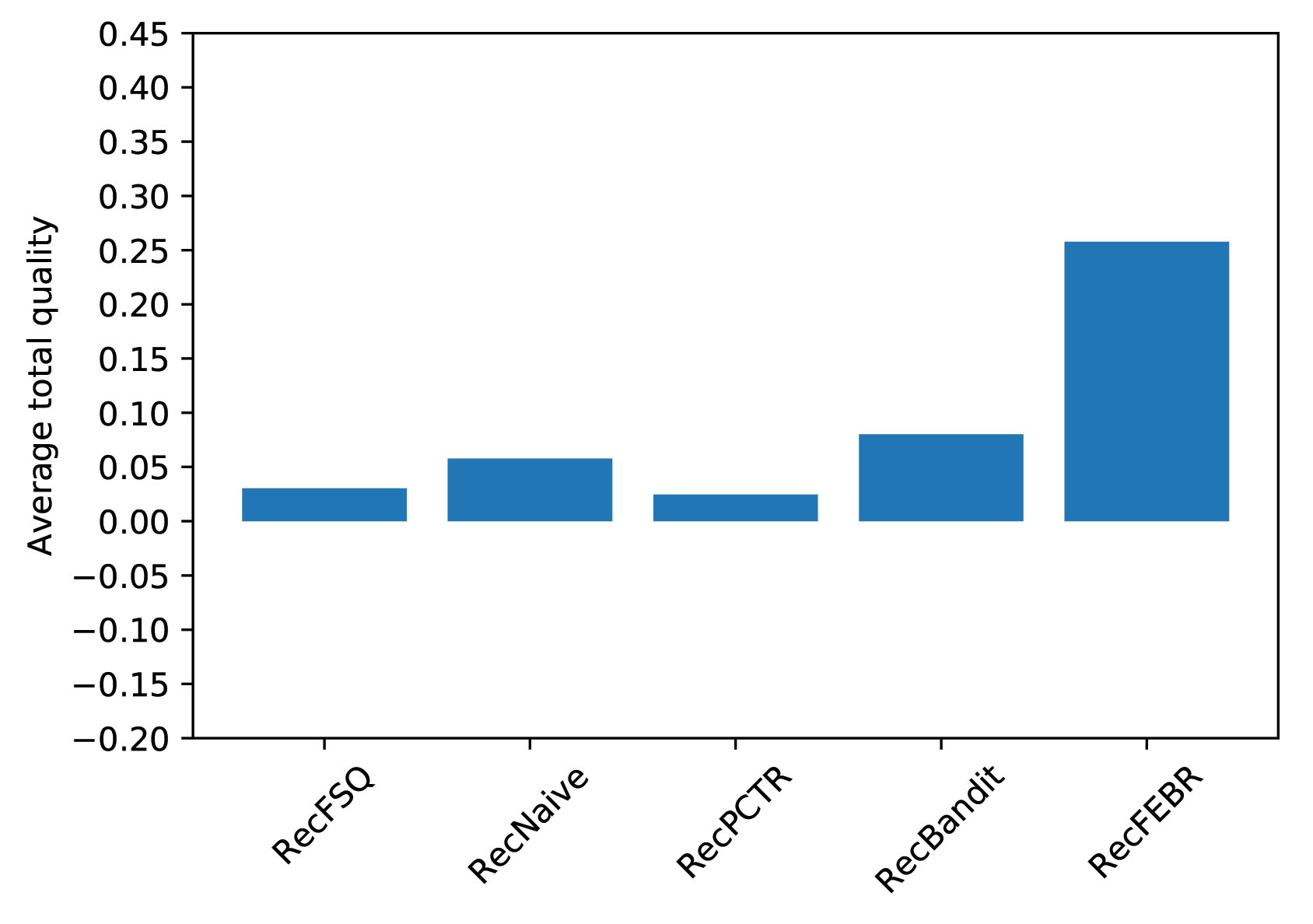}
\caption{Comparison of the average total quality $Q_T$ of baseline approaches and \textit{RecFEBR}.}
\label{fig:ttq}
\end{minipage}
\end{figure}

In Figure~\ref{fig:ttq}, we notice that our approach provides the highest value of $Q_T$  compared to the baseline approaches, that achieve lower quality values. In addition, Figure~\ref{fig:wtime} shows that this significant gain in quality does not come at the cost of losing user engagement, since RecFEBR reaches almost the same total watch time as the best ones (\textit{RecFSQ} and \textit{RecBandit}). Thus, we can confirm the three following points.

First, FEBR can effectively guide users through positive quality content by proposing relevant and beneficial recommendations, matching their preferences and tastes.
The approach generally succeeds (through its AL/IRL component) to learn an optimal expert policy $\pi^*_e$.
It also manages (through its recommendation strategy) to take advantage of $\pi^*_e$, and to associate a considerable number of user states  to their closest similar expert states, based on their common observable features using the proposed classification model.
Thus, regardless of the values of $Q_T$, which highly depends on the models used by the simulation environments, we can conclude that this simulation instance proves the efficiency of our approach in adapting IRL to capture an expert behavior (despite the complexity and the ambiguity it involves) for personalized recommendations, by ensuring a significant (positive) impact on $Q_T$.
Moreover, we highlight the important role of the classification model, that is: an accurate model trained on large datasets (built by a large number of experts and containing high number of states) could be more efficient, and significantly improve performances.

Second, although \textit{RecNaive} recommends evaluated videos, this approach will require an important number of evaluated videos (at the scale of the system size) to be injected to (hopefully) improve the quality of watched videos. However, even if that happens (which is very unlikely in large systems like YouTube), the random agent may fail (or likely take a long time) to learn a policy capable of delivering high-quality and personalized content. Then, this naive exploitation of these videos, making continuous explorations by the agent (because of random actions) will often recommend irrelevant contents. Thus, regardless of the higher quality that these recommendations can ensure in some successful scenarios, users will often skip recommended videos and interrupt watching, as shown by the value of $W_T$ in Figure~\ref{fig:wtime}.

Third, as we can see in Figure \ref{fig:ttq}, optimizing the quality using \textit{RecFSQ}, \textit{RecPCTR} and \textit{RecBandit}  cannot ensure a sufficient exposure to content of high quality. The low values of $Q_T$ (compared to $W_T$) achieved by these RL methods can be explained by the ability of the RS to excessively act under user objectives. This behavior (contrary to what may seem desirable) could result in unprofitable actions  (or even worse)  by the RL agent, again and again. This may happen because the state quality is measured based on system engagement metrics, which incentivizes agents to prefer personalization over high quality of recommendation. In this context, Figure \ref{fig:ttq} shows the good personalization results of these methods, except for \textit{RecPCTR}, which produces a relatively low value of $W_T$. The result of \textit{RecPCTR} regarding $W_T$ could be explained by its myopic nature.

\section{Conclusion}
\label{sec_conc}

In this paper, we proposed FEBR, a configurable AL/IRL-based framework for quality content recommendations leveraging expert evaluations. We developed a MDP environment that exploits the experts' knowledge and their personal preferences to derive an optimal policy that maximizes an unknown reward function, using the MaxEnt IRL model. We then used this policy to build our recommender agent (the classification model), which matches user and expert's states and recommends the best related action. Experiments on video recommendations (using a simulated RL-based RS) show that expert-driven recommendations using our approach could be an efficient solution to control the quality of the recommended content, while keeping a high user engagement rate. 

For future works, an important challenge would be to generalize this approach to systems of large dimension. Besides, one could extend this work to study the case of unreliable experts who tend to manipulate the system for malicious or personal purposes. It is also interesting to mitigate the effects of the cold start problem in the end-user recommendation process, which indirectly affects recommendations in situation of poor state matching by the classifier (which may result in beneficial but non-personalized recommendations). It may also be interesting to develop a decentralized version of the framework using some deep neural networks approaches to better learn the expert's policy (with less constraints on the reward function), enabling a better scalability for systems of high dimensions, with less constraints on the reward functions. It may also bring more insights for running similar experiments in real-life conditions.

\newpage 

\bibliographystyle{ACM-Reference-Format}
\bibliography{febr_main}

\end{document}